\documentstyle{article} 

\newtheorem{prop}{Proposition}[section]
\newtheorem{defn}{Definition}[section]

\begin{document}

%\normalbaselineskip=2.0\normalbaselineskip
%\normalbaselines

\centerline{\bf{On the Geometry of Certain Isospectral Sets}}
\centerline{\bf{in the Full Kostant-Toda Lattice}}

\bigskip

\centerline{Barbara Shipman}

\bigskip

\centerline{\it Department of Mathematics, University of Rochester}
\centerline{\it Rochester, NY 14627, USA}  
\centerline{e-mail: shipman@cauchy.math.rochester.edu}

\bigskip

\noindent {\bf Abstract:}
We use momentum mappings on generalized flag manifolds and their momentum
polytopes to study the geometry of the level sets of the
1-chop integrals of the full Kostant-Toda lattice in certain
isospectral submanifolds of the phase space.  We derive expressions
for these integrals in terms of Pl\"ucker coordinates on the flag
manifold in the case that all eigenvalues are zero and compare the
geometry of the base locus of their level set varieties with the
corresponding geometry for distinct eigenvalues.
Finally, we illustrate and extend our results in the context of the
full $sl(3,{\bf C})$ and $sl(4,{\bf C})$ Kostant-Toda lattices.

\section{Introduction}

The system of differential equations for the full Kostant-Toda lattice
is written in Lax form as
%The full Kostant-Toda lattice is the system whose evolution is
%governed by the Toda equations written in Lax form as
\begin{equation}
\dot{X}(t) = [X(t), \Pi_{{\cal N}_-} X(t)], \label{Toda}
\end{equation}
in which $X$ belongs to the space of matrices
$$ \epsilon + {\cal B}_- = \Bigg\{ \left(
\begin{array}{ccccc} 
* & 1 & 0 & \cdots & 0 \\ 
\vdots & \ddots & \ddots & \ddots & \vdots \\
\vdots & \ddots & \ddots & \ddots & 0 \\
\vdots & \ddots & \ddots & \ddots & 1 \\
* & \cdots & \cdots & \cdots & *
\end{array} \right) : trX = 0  \Bigg\}, 
$$
where ${\cal B}_-$ is the Lie algebra of lower triangular matrices
with trace zero, $\epsilon$ is the matrix with 1's on the
superdiagonal and zeros elsewhere, and $\Pi_{{\cal N}_-} X$ is the
strictly lower triangular part of $X$.  The space $\epsilon + {\cal
B}_-$ is not a symplectic manifold.  It is a Poisson manifold which is
foliated by symplectic leaves of different dimensions.  The full
Kostant-Toda lattice is known to be completely integrable 
(\cite{DLNT,Sing,EFS}) on its generic symplectic leaves; moreover, it
is a system 
for which there exist distinct maximal involutive families of
integrals.   

In \cite{EFS} the authors consider a particular family of integrals
known as the 
``k-chop integrals'' for $k = 0, \ldots, [{n-1 \over 2}]$, which we 
describe in Section 2.  The 0-chop integrals are simply the
eigenvalues of $X \in \epsilon + {\cal B}_-$.  Indeed, the form of the
solution readily shows that these are constants of motion.
To solve the system (\ref{Toda}) with initial condition $X_0$, we factor the
exponential $e^{t X_0}$ as
$$e^{t X_0} = n(t) b(t),$$
where $n(t)$ is lower unipotent and $b(t)$ is upper triangular.  
Conjugating $X_0$ by $n(t)$ yields the solution, $X(t)$ (\cite{R,RS}):  
\begin{equation} X(t) = n^{-1}(t) \ X_0 \ n(t), \label{solution} 
\end{equation}
which clearly preserves the spectrum.  We point out here that
the k-chop integrals for $k > 0$ are rational functions in the entries
of $X$ and that typically the flows associated to these integrals exit
the phase space in finite (complex) time.  In particular,
if this factorization cannot be done for some $t =
t_0$, then $X(t)$ has a pole at $t_0$.  

The geometry of the flag manifold
$Sl(n,{\bf C})/B$, where $B$ is the subgroup of upper triangular matrices,
is used in \cite{EFS}
to study the restriction of (\ref{Toda}) to an isospectral submanifold
of $\epsilon +  
{\cal B}_-$ in which the eigenvalues are distinct.  Using a
factorization result of Kostant, one can define an embedding of this
submanifold into $Sl(n,{\bf C})/B$ as an open dense set. 
Under this embedding, the Toda flows associated to the
0-chop integrals generate the action of the diagonal complex torus
$({\bf C}^*)^{n-1}$ on the flag manifold so that the compactified level sets
of the constants of motion are unions of torus orbits.  This embedding
has therefore become known as the ``torus embedding.''  

The fact that these compactified level sets are invariant under the
action of the diagonal torus is the basis of the study of
the full Kostant-Toda lattice in \cite{diss}, which brings in the geometry of
certain convex polytopes associated to these torus orbits through the
momentum mapping, extending some of the results in \cite{EFS} and
providing geometric descriptions of certain distinctive features of the
system including the symplectic leaf stratification, the relationship
between level sets cut out by different involutive families of
integrals, and monodromy near certain ``degenerate'' level sets.

This paper continues in the same spirit, appealing to the structure of
momentum polytopes to illuminate the geometry of the full Kostant-Toda
lattice.    
In Section 4 we describe in a more general setting the momentum mapping on
the flag manifold associated to the action of the diagonal compact
torus, $(S^1)^{n-1}$, and a convexity theorem which describes the
convex polytopes arising as the images under this mapping of the
closures of the orbits of $({\bf C}^*)^{n-1}$.

It is shown in \cite{EFS} that under the torus embedding, the 1-chop
integrals have simple expressions in terms of certain
homogeneous coordinates $[\pi_i] \times [\pi_i^*]$ on $P^n \times
(P^n)^*$.  For example, when $n = 3$, the single 1-chop integral has
the form
$$\frac{\lambda_2 \lambda_3 \pi_1 \pi_1^* + \lambda_1 \lambda_3 \pi_2
\pi_2^* + \lambda_1 \lambda_2 \pi_3 \pi_3^*}{\lambda_1 \pi_1 \pi_1^* +
\lambda_2 \pi_2 \pi_2^* + \lambda_3 \pi_3 \pi_3^*},$$
which clearly shows its invariance under the torus $({\bf C}^*)^2$ since
the coordinates $\pi_i$ and $\pi^*_i$ are scaled under the flow
in such a way that the products $\pi_i \pi_i^*$ remain unchanged.
Moreover, the 1-chop integrals for general $n$ depend only on the 
projection of the flag manifold to the manifold of partial flags $\{
V^1 \subset V^{n-1} \subset {\bf C}^n \} \subset P^n \times (P^n)^*$, which
we denote by $Sl(n,{\bf C})/P_1$. Clearing the denominators in the
expressions of these integrals defines a collection of subvarieties in
this partial flag manifold.  We show that the
intersection of these varieties, their base locus, has a simple
description in terms of the momentum mapping on $Sl(n,{\bf C})/P_1$ and its
momentum polytope.  Indeed, this base locus is precisely the inverse
image under the momentum mapping of the boundary of the momentum
polytope. 

Our discussion so far and the treatments of the full Kostant-Toda
lattice in \cite{EFS} and \cite{diss} concern only the situation in
which the eigenvalues of $X \in \epsilon + {\cal B}_-$ are distinct.
One of the reasons for this is that it is only in this case that the
torus embedding is defined, producing compactified level sets of the
constants of motion which are unions of orbits of the diagonal torus.
There is, however, a different mapping which embeds an arbitrary
isospectral submanifold of $\epsilon + {\cal B}_-$ into the flag
manifold.  In this case the Toda flows for the 0-chop integrals
generate the action (by left multiplication) of the exponential of the
abelian nilpotent algebra whose elements have the form
$$\left( \begin{array}{cccc} 0 & a_1 & \cdots & a_{n-1} \\
\vdots & \ddots & \ddots & \vdots \\ \vdots & \ddots & \ddots & a_1 \\
0 & \cdots & \cdots & 0 \end{array} \right),$$ 
where the $a_i$ belong to ${\bf C}$, on the flag manifold.  This embedding
is called the ``companion embedding'' since the role of
the diagonal matrix in the torus embedding is replaced by the
companion matrix of $X \in \epsilon + {\cal B}_-$.  

Computing the expressions for the 1-chop integrals in
terms of the Pl\"ucker coordinates on the flag manifold under the
companion embedding, we discover that as in the case of the torus
embedding, they involve only the coordinates $[\pi_i] \times
[\pi_i^*]$ on the partial flag manifold $Sl(n,{\bf C})/P_1$.
So as in the case of the torus embedding when the eigenvalues are
distinct, we obtain again a collection of subvarieties of $Sl(n,{\bf C})/P_1$.

However, since the varieties in $Sl(n,{\bf C})/P_1$ defined
by the 1-chop integrals in the case of the companion embedding are not
preserved by the diagonal 
torus, the level sets of the constants of motion are not unions of
orbits of this torus and we cannot apply the convexity theorem
directly to study their geometry.  Neither does the common
intersection (base locus) of these varieties for an arbitrary choice of
isospectral level set lend itself to a geometric description in terms
of momentum polytopes.  However, in the case of the isospectral
submanifold $(\epsilon + {\cal B}_-)_0$, in which all the eigenvalues
coincide and are equal to zero, the base locus of the varieties
determined by the 1-chop integrals is
invariant under the action of the diagonal torus.  Here the companion
matrix is simply the nilpotent matrix with 1's on the superdiagonal
and 0's elsewhere, and the expressions of the 1-chop integrals become
much simpler.
When $n = 3$, for example, the single 1-chop integral is
$$\frac{\pi_3 \pi_1^*}{\pi_2 \pi_1^* + \pi_3 \pi_2^*}.$$

In this paper we refer to the situation in which an isospectral
submanifold of $\epsilon + {\cal B}_-$ with distinct eigenvalues is
embedded into the flag manifold by the torus embedding as Case A and
to the companion embedding of $(\epsilon + {\cal B}_-)_0$ into
$Sl(n,{\bf C})/B$ as Case B.  We show that in Case B the base locus of the
varieties in $Sl(n,{\bf {\bf C}})/P_1$ determined by the 1-chop integrals 
is not the complete inverse image under the
momentum mapping of the boundary of the momentum polytope, in
contrast to what we find  
in Case A.  Here the base locus corresponds to only certain faces of
this polytope, one associated to each of the fundamental weights of
$sl(n,\bf C)$.  

It is interesting to note that first, it is in the two extreme cases,
the generic case in which all eigenvalues are distinct and
the most degenerate case in which all eigenvalues coincide, that the
base locus of the varieties in $Sl(n,{\bf C})/P_1$ corresponding to the
1-chop integrals (under the appropriate embedding) have simple
descriptions in terms of the momentum polytope, and secondly that the
two embeddings that reveal this geometry in the two
cases are different.  

After first developing our results for Cases A and B in Section 4, we
consider the expressions of the 1-chop integrals for an arbitrary
isospectral level set under the consistent use of the companion
embedding.  The expressions 
we obtain provide some insight into how the 1-chop integrals under
this embedding simplify as a level set with distinct eigenvalues
degenerates to the extreme case in which all
eigenvalues coincide (at zero).

We devote the final two sections 
to the special cases $n=3$ and $n=4$.  For $n=3$ we consider in more
detail the geometry of the level sets of the single 1-chop integral (a
Casimir) in the flag manifold in both Case A and Case B and show how
the common intersection of the level set varieties in each of these cases
precisely encodes the stratification of the relevant isospectral
submanifold  induced by the symplectic stratification of 
$\epsilon + {\cal B}_-$.  In our example with $n=4$, after
illustrating our results for the 1-chop integrals, we consider a
different maximal involutive family of integrals whose geometry is
studied in detail in \cite{diss}.  We obtain for this family of
integrals statements which are completely analogous to those for the
1-chop integrals, the 
partial flag manifold $Sl(4,{\bf C})/P_1$ being replaced by the Grassmannian
$G(2,4)$ of two-dimensional subspaces of ${\bf C}^4$.

\section{Background}

To describe the Poisson structure on $\epsilon + {\cal B}_-$ we
consider the decomposition of the Lie 
algebra $sl(n,{\bf C})$ into its upper triangular and lower nilpotent
subalgebras:  
$$sl(n,{\bf C}) = {\cal B}_+ \oplus {\cal N}_- .$$
Using the nondegenerate Killing form $<X,Y> = 2n \cdot tr(XY)$ on
$sl(n,{\bf C})$, 
we identify the dual, ${\cal B}_+^*$, of ${\cal B}_+$ with ${\cal
N}_-^{\perp}$, which is ${\cal B}_-$.  This gives an
identification of $\epsilon + {\cal B}_-$ with the dual of a Lie
algebra:
$$\epsilon + {\cal B}_- \, \cong \, {\cal B}_- \,  = \,  {\cal
N}_-^{\perp}\,  \cong \,  {\cal B}_+^*.$$ 
The space $\epsilon + {\cal B}_-$ acquires its Poisson structure
through this 
identification from the Lie-Poisson structure on ${\cal B}_+^*$, whose
symplectic leaves are the coadjoint orbits.  
Indeed, the differential equations
generated by the Hamiltonian $H = {1 \over 2} tr X^2$
are precisely the Toda equations (\ref{Toda}).  

From the solution (\ref{solution}), we know that the functions $I_{k0} =
{1 \over k} tr X^k$ for $k = 2, \ldots, n$ are constants of motion.
To obtain a completely integrable system on a generic symplectic leaf,
one must find Casimirs and sufficiently many additional integrals.  
The full Kostant-Toda lattice is an integrable system
for which there exist distinct maximal families of constants of motion in
involution, where the constants in  different families do not
necessarily commute with each other.  Using a method introduced by
Thimm \cite{Th} and applied to the Toda lattice in \cite{DLNT}, such involutive
families may be found by considering different nested chains of
parabolic subalgebras of $sl(n,{\bf C})$.  \cite{EFS} deals primarily with a
particular family of integrals.

\begin{prop} \label{k-chopdefn} (\cite{DLNT,EFS}) 
For $k = 0, \ldots, [{(n-1) \over 2}]$, denote by $(X - \lambda
Id)_{(k)}$ the result of removing the first $k$ rows and last $k$
columns from $X - \lambda Id$, and let $\lambda_{rk}$, $r = 1,
\ldots, n - 2k$, denote the roots of
$$\tilde Q_k(X,\lambda) = det(X - \lambda Id)_{(k)} = E_{0k}
\lambda^{n - 2k} + \cdots + E_{n-2k,k}.$$
The $\lambda_{rk}$ are constants of motion for the full Kostant-Toda
lattice.   The coefficients, $I_{rk}$, of the monic polynomial
$$ Q_k(X,\lambda) = {det(X - \lambda Id)_{(k)} \over E_{0k}} = 
\lambda^{n-2k} + I_{1k} \lambda^{n-2k-1} + \cdots + I_{n-2k,k}$$
are constants of motion equivalent to the $\lambda_{rk}$.  They are
called the $k$-chop integrals.  (The $I_{r0}$ are the
coefficients of the characteristic polynomial of $X$.)  The functions
$I_{1k} = \sum_r \lambda_{rk}$ are  Casimirs on $\epsilon + {\cal
B}_-$, and the $I_{rk}$ for $r > 1$ constitute a complete involutive
family of integrals for the generic symplectic leaves of $\epsilon +
{\cal B}_-$ cut out by the Casimirs $I_{1k}$. 
\end{prop}

In \cite{EFS}, the authors present a method of computing the $k$-chop
integrals.

\begin{prop} \label{k-chopcalc} (\cite{EFS})
Choose $X \in \epsilon + {\cal B}_-$, and break it into blocks of the
indicated sizes, 
$$ X = \bordermatrix{&  k & n-2k & k \cr k & X_1 & X_2 & X_3 \cr
n-2k & X_4 & X_5 & X_6 \cr k & X_7 & X_8 & X_9 \cr}, $$
 where $k$ is an integer, $0 \leq k \leq [{(n-1) \over
2}]$.  If $det X_7 \neq 0$, define the matrix $\phi_k(X)$ by
\begin{eqnarray*} \phi_k(X) & = & X_5 - X_4 X_7^{-1} X_8 \in 
Gl(n-2k,{\bf C}), \quad k \neq 0, \\
\phi_0(X) & = & X. 
\end{eqnarray*}
Then the $I_{rk}$ are the coefficients of the polynomial $det(\lambda
- \phi_k(X))$.
\end{prop}

The  ``generic'' level sets of these
constants of motion are studied in \cite{EFS} in terms of the geometry
of generalized flag 
manifolds.  This is made possible through the following result of
Kostant.
 
\begin{prop} \label{normalform} (\cite{K})
Let $\lambda^n - s_2 \lambda^{n-2} - \ldots - s_n$ be the
characteristic polynomial of $X \in \epsilon + {\cal B}_-$, and let 
$$C = \left( \begin{array}{ccccc} 0 & 1 & 0 & \cdots & 0 \\
\vdots & \ddots & \ddots & \ddots & \vdots \\
\vdots & \ddots & \ddots & \ddots & 0 \\
0 & \cdots & \cdots & \ddots & 1 \\
s_n & \cdots & \cdots & s_2 & 0  \end{array}  \right) $$
be the companion matrix.  Then there is a unique lower unipotent
matrix $L$ such that $X = L C L^{-1}$.
\end{prop}

Now fix the eigenvalues $\lambda_1, \ldots, \lambda_n$ of $X \in
\epsilon + {\cal B}_-$.  Let $\Lambda$ be the diagonal matrix 
$\Lambda = \mbox{diag}(\lambda_1, \ldots, \lambda_n)$
for some fixed ordering of the $\lambda_i$, and denote by 
$(\epsilon + {\cal B}_-)_{\Lambda}$ the isospectral submanifold of
$\epsilon + {\cal B}_-$ with these eigenvalues. 
Because of the uniqueness of $L$, there is an embedding of $(\epsilon +
{\cal B}_-)_{\Lambda}$ into the flag manifold $Sl(n,{\bf C})/B$, where
$B$ is the   
subgroup of upper triangular matrices, defined by 
$$\Phi_{\Lambda} : (\epsilon + {\cal B}_-)_{\Lambda} \rightarrow
Sl(n,{\bf C})/B$$ 
$$X \mapsto L^{-1} \bmod B.$$
This is known as the {\it companion embedding}.

Now suppose that 
the eigenvalues $\lambda_i$ are distinct. Then $C = V \Lambda
V^{-1}$, where $V$ is a Vandermonde matrix, and the theorem implies
that for each $X$ in $(\epsilon~+~{\cal
B}_-)_{\Lambda}$, there is a unique lower unipotent matrix $L$ such
that $X = L V \Lambda V^{-1} L^{-1}$. Again, the uniqueness of $L$ implies
that the mapping 
$$\Psi_{\Lambda} : (\epsilon + {\cal B}_-)_{\Lambda} \rightarrow
Sl(n,{\bf C})/B$$ 
$$X \mapsto V^{-1} L^{-1} \bmod B$$
is also an embedding.

Since the flows for the integrals ${1 \over k} tr X^k$ generate the
action of the diagonal complex torus $({\bf C}^*)^{n-1}$ on the flag
manifold under this 
embedding (see \cite{EFS}), it is called the {\it torus embedding}.
Under this mapping, each compactified level set of the constants of
motion is a union of torus orbits in the flag
manifold.  By an application of a more general convexity theorem of
Atiyah in \cite{AtiyahCCH}, we can associate to the closure of each torus
orbit a convex polytope by means of the momentum mapping, as described
in the next section. The geometry of generalized flag manifolds and
their momentum 
polytopes is the basis of the study of the full Kostant-Toda lattice
in \cite{diss}, in which the eigenvalues are taken to be distinct.  In
this paper we extend some of the work in \cite{diss} to the case in
which all the eigenvalues are equal to zero and compare our results to
the corresponding statements in the case of distinct eigenvalues.

\section{The Momentum Mapping on $G/P$}

In this section we first introduce the momentum mapping and discuss
the convexity 
theorem in the more general setting in which $Sl(n,{\bf C})$ is replaced
by an arbitrary complex semisimple Lie group.  We then specialize our
discussion to the case of $Sl(n,{\bf C})$ and establish some notation.

Let $G$ be a complex semisimple Lie group, $H$ a Cartan subgroup of
$G$, and $B$ a Borel subgroup containing $H$.  We denote by $\cal H$
the Lie algebra of $H$ and by ${\cal H}^*$ the dual of $\cal H$.  Let
$P$ be a parabolic subgroup of $G$ containing $B$.  The homogeneous
space $G/P$ can be realized as the orbit of $G$ through a
projectivized highest weight vector in the projectivization, $P(V)$, of
an irreducible representation $\rho : G \rightarrow Gl(V)$
(see \cite{GelfandSerganova}, for example).

Because $G/P$ is a projective algebraic variety, it is
K\"ahler and is therefore also a symplectic manifold.  The
action of the compact torus $T$ in $H$ preserves the symplectic
structure.  This gives rise to a momentum mapping $\mu : G/P
\rightarrow \tau^*$, where $\tau^*$ is the dual of the Lie algebra of
$T$.  ($\tau^*$ is the real part of ${\cal H}^*$.)  To describe this
mapping, let $\cal A$ be the set of weights of $\rho$ taken with
multiplicity, and choose a basis of weight vectors $\{ v_{\alpha} :
\alpha \in \cal A \}$ for $V$.  A point $[X] \in G/P \subset P(V)$,
represented by a vector $X \in V$, determines uniquely up to a scalar
the collection of numbers $\pi_{\alpha}(X)$, where $X = \sum_{\alpha
\in \cal A} \pi_{\alpha}(X) v_{\alpha}$.  We refer to the homogeneous
coordinates $[\pi_{\alpha}(X)]$ on $G/P$ as {\it Pl\"ucker
coordinates}. 

\bigskip

\bigskip

\noindent The momentum mapping for the
action of $T$ on $G/P$ is defined by
$$\mu : G/P \rightarrow \tau^*$$
$$[X] \mapsto \frac{\sum_{\alpha \in \cal A} |\pi_{\alpha}(X)|^2
\alpha}{\sum_{\alpha \in \cal A} |\pi_{\alpha}(X)|^2}.$$

Let $H \cdot [X]$ be the orbit of $H$ through $[X]$, and let
$\overline{H \cdot [X]}$ be its closure in $G/P$.  To describe the
image of $\overline{H \cdot [X]}$ under the momentum mapping, we
consider the orbit $W \cdot \alpha^V$ of the Weyl group $W$ of $G$
through the highest weight vector $\alpha^V$ of $V$ determined by our
choice of $B$.  It is shown in \cite{GelfandSerganova} using the more
general results in \cite{AtiyahCCH} that $\mu(\overline{H \cdot 
[X]})$ is the convex polytope whose vertices are the weights 
$\{\alpha \in W \cdot \alpha^V : \pi_{\alpha}(X) \neq 0 \}$.  In
particular, the image of $G/P$ under the momentum mapping is the
convex hull of the orbit of the Weyl group through $\alpha^V$.  This
image is therefore the weight polytope of the irreducible
representation $V$; we refer to it also as the {\it momentum polytope}
of $G/P$.  The vertices of this polytope are the images under $\mu$ of
the fixed points, $[v_{\alpha}]$ for $\alpha \in W \cdot \alpha^V$, of
the torus action; $\mu([v_{\alpha}]) = \alpha$.

The momentum mapping induces a partition of $G/P$ into equivalence
classes called {\it strata} (see \cite{GelfandSerganova}), where each
stratum is the union of all torus orbits in $G/P$ whose closures have
as their images under the momentum mapping the same convex polytope.
The generic stratum is the union of all torus orbits whose closures
have the full momentum polytope as their common image under $\mu$.
These are the 
generic torus orbits, on which no Pl\"ucker coordinate vanishes.

Now let $G$ be the group $Sl(n,{\bf C})$, $H$ its diagonal subgroup, and $V$
the adjoint representation of $sl(n,{\bf C})$.  We denote by $L_i$ the
linear function in ${\cal H}^*$ which sends an element of $\cal H$ to
its $i^{th}$ diagonal entry.  Then ${\cal H}^*$ is the quotient space
$${\cal H}^* = \{ \sum_{i = 1}^n c_i L_i : c_i \in {\bf C} \} / < \sum_{i =
1}^n L_i > \, \, \cong {\bf C}^{n-1},$$
and the roots of the adjoint representation are $L_i - L_j$, $i \neq
j$.  These roots are the vertices of the weight polytope of $V$,
which we denote by $\triangle_n$; they constitute the orbit of the
Weyl group of $Sl(n,{\bf C})$, the symmetric group on $n$ elements, through
the highest weight $L_1 - L_n$.  The adjoint representation of
$sl(n,{\bf C})$ is the $(n^2 - 1)$-dimensional irreducible component in the
representation ${\bf C}^n \otimes \wedge^{n-1} {\bf C}^n$, where ${\bf
C}^n$ is the 
standard representation of $sl(n,{\bf C})$.  

Let $\{ e_i \}_{i = 1}^n$ be
the standard basis of ${\bf C}^n$.  We identify $\wedge^{n-1} {\bf
C}^n$ with the 
dual of ${\bf C}^n$ by requiring that $\omega(\beta)$ for $\omega \in 
\wedge^{n-1} {\bf C}^n$ and $\beta \in {\bf C}^n$ satisfy
$\beta \wedge \omega = \omega(\beta)e_1 \wedge \ldots \wedge e_n$.
Under this identification, the dual basis of $\{ e_i \}_{i =
1}^n$ is $\{ e_i^* \}_{i = 1}^n$, where
$e_i^* = (-1)^{i + 1} e_1 \wedge \ldots \wedge e_{i - 1} \wedge e_{i +
1} \ldots \wedge e_n$.  Our Pl\"ucker coordinates on $G \cdot 
[e_1 \otimes e_n^*] \subset P(V)$ are defined with respect to the
weight vectors $v_{L_i - 
L_j} = e_i \otimes e_j^*$ for the root spaces of $sl(n,{\bf C})$  and the
basis $\{e_i \otimes e_i^* - e_{i + 1} \otimes e_{i + 1}^*\}_{i =
1}^{n-1}$ of $\cal H$.  (The vectors $e_i
\otimes e_i^*$, $i = 1, \ldots, n$ span the $n$-dimensional zero
weight space in the tensor product ${\bf C}^n \otimes \wedge^{n-1} {\bf C}^n$.)

The orbit $G \cdot [e_1 \otimes e_n^*]$ is realized as the homogeneous space 
$G/P_1$, where $P_1$ is the stabilizer of $[e_1 \otimes e_n^*]$ in
$P(V)$. It
consists of the elements $[v \otimes \sigma] \in P~({\bf
C}^n~\otimes~\wedge^{n-1}~{\bf C}^n)$ for which $v \wedge \sigma = 0$.
Its obvious 
embedding into 
$P({\bf C}^n) \times  P(\wedge^{n-1} {\bf C}^n)$ defines coordinates $[\pi_1 :
\cdots : 
\pi_n] \times [\pi_1^* : \cdots : \pi_n^*]$ on $G/P_1$, where 
$[\pi_i]$ are the Pl\"ucker coordinates on
$P({\bf C}^n)$ with respect to the standard basis and $[\pi_i^*]$ are 
the Pl\"ucker coordinates on
$P(\wedge^{n-1} {\bf C}^n)$ with respect to the dual basis $\{e_i^*\}$.  
(If we represent $[X] \in G/P_1$ by $M$ mod $P_1$ with $M \in G$, then
$[\pi_i] = [M_{i1}]$, and $[\pi_i^*] = [(-1)^{i+1}M_{\hat i \hat n}]$,
where $M_{\hat i \hat n}$ is the determinant of the matrix obtained by
deleting the $i^{th}$ row and the $n^{th}$ column of $M$.)  In these
coordinates, the variety $G/P_1 \subset P({\bf C}^n) \times
P(\wedge^{n-1} {\bf C}^n)$ is cut out by the equation $\sum_{i = 1}^n \pi_i
\pi_i^* = 0$.  

It will be convenient to use both the Pl\"ucker
coordinates $[\pi_{\alpha}]$ for the embedding $G/P_1 \rightarrow
P(V)$ and the coordinates $[\pi_i] \times [\pi_i^*]$ for the embedding 
$G/P_1 \rightarrow P({\bf C}^n) \times P(\wedge^{n-1} {\bf C}^n)$.
For $[X] \in 
G/P_1$, $X$ has the form
$$X = \sum_{i \neq j} \pi_{L_i - L_j}(X) e_i \otimes e_j^* +
\sum_{i=1}^n a_i(X) e_i \otimes e_i^*,$$
where $\sum_{i=1}^n a_i(X) = 0$.  In terms of the coordinates 
$[\pi_i] \times [\pi_i^*]$ on $[X]$, the Pl\"ucker coordinates
$\pi_{L_i - L_j}(X)$ are 
projectively equal to the products $\pi_i \pi_j^*$ for $i \neq j$ :
$$[\pi_{L_i - L_j}(X)]_{i \neq j} = [\pi_i \pi_j^*]_{i \neq j},$$
and the $a_i(X)$ are projectively equal to the products $\pi_i
\pi_i^*$  : 
$$[a_i(X)]_{i=1}^n = [\pi_i \pi_i^*]_{i=1}^n.$$
 
Throughout the paper, when we refer to a torus orbit, we mean an orbit
of the diagonal complex torus $({\bf C}^*)^{(n-1)}$, and reference to a face
or edge of a polytope means its closure.

\section{Results for General $Sl(n,{\bf C})$}

We consider two types of isospectral submanifolds of $\epsilon + {\cal
B}_-$, those with distinct eigenvalues and the one whose eigenvalues
are all equal to zero, which we denote by $(\epsilon + {\cal
B}_-)_0$.  In the case that the eigenvalues are distinct, we take the
torus embedding of $(\epsilon + {\cal B}_-)_{\Lambda}$ into $G/B$; we
embed $(\epsilon + {\cal B}_-)_0$ into the flag manifold by the
companion embedding.  We refer to these as Case A and Case B,
respectively.  

\subsection{The two extreme cases}

It is shown in \cite{EFS} that under the torus embedding of 
$(\epsilon + {\cal B}_-)_{\Lambda}$ (with distinct eigenvalues) into
$G/B$, the 1-chop integrals have the expressions
$$I_{r1} = \frac{\sum_{i=1}^n \sigma_{r+1}(\hat i) \pi_i \pi_i^*}
{\sum_{i=1}^n \sigma_{1}(\hat i) \pi_i \pi_i^*},$$
where $\sigma_j (\hat i)$ is the $j^{th}$ symmetric polynomial on the
$n-1$ eigenvalues different from $\lambda_i$.  Observe that these
functions on $G/B$ involve only the Pl\"ucker coordinates $[\pi_i]
\times [\pi_i^*]$ and are therefore defined in terms of the projection
of the flag manifold to $G/P_1$.  The situation in Case B is similar.

\begin{prop} \label{BIr1expr} Under the companion embedding of
$(\epsilon + {\cal 
B}_-)_0$ into the flag manifold, the 1-chop integrals have the
expressions
$$I_{r1} = \frac{\sum_{i=1}^{n-r-1} \pi_{i+r+1} \pi_i^*}
{\sum_{i=1}^{n-1}\pi_{i+1} \pi_i^*},$$
which depend only on the projection of $G/B$ to
$G/P_1$.  In terms of the Pl\"ucker coordinates $\pi_{\alpha}$ on
$G/P_1$, these expressions are
$$I_{r1} = \frac{\sum_{ht(\alpha) = -1-r}
\pi_{\alpha}}{\sum_{ht(\alpha) = -1} \pi_{\alpha}},$$
where $ht(\alpha)$ is the height of the weight $\alpha$.
\end{prop}

\noindent{\bf Proof.}  Let $C_0$ denote the nilpotent matrix with 1's
on the superdiagonal and zeros elsewhere, and for $M \in Sl(n,{\bf C})$,
denote by $M_{\hat i 
\hat j}$ the determinant of the matrix obtained by deleting row $i$
and column $j$ from $M$.  Then for $X \in (\epsilon + {\cal
B}_-)_0$,
\begin{eqnarray*}
det(X - \lambda I)_{(1)} & = & (L(C_0 - \lambda I) L^{-1})_{\hat 1
\hat n} \\
& = & \sum_{i,j = 1}^n L_{\hat 1 \hat i}(C_0 - \lambda I)_{\hat i \hat
j} L^{-1}_{\hat j \hat n} \\
& = & \sum_{i,j = 1}^n (-1)^{i+1} L^{-1}_{i1} L^{-1}_{\hat j \hat
n}(C_0 - \lambda 
I)_{\hat i \hat j} \\
& = & \sum_{i,j = 1}^n (-1)^{i+1} \pi_i (-1)^{j+1} \pi_j^* (C_0 - \lambda
I)_{\hat i \hat j} \\
& = & \sum_{i=1}^n \sum_{1 \leq j \leq i} \pi_i \pi_j^*
\lambda^{n-1-(i-j)},
\end{eqnarray*}
where $\pi_i$ and $\pi_i^*$ denote $\pi_i(\Phi_0(X))$ and
$\pi_i^*(\Phi_0(X))$, respectively.  By Proposition~\ref{k-chopdefn},
$I_{r1}$ is 
the coefficient of $\lambda^{n-r-2}$ in this polynomial divided by the
coefficient of $\lambda^{n-2}$.  (The coefficient of $\lambda^{n-1}$
is the Pl\"ucker relation $\sum_{i=1}^n \pi_i \pi_i^*$, which vanishes.)
\rule[-1mm]{1mm}{2mm}

\bigskip

\noindent {\bf Remark:}  Observe that in Case A, the expressions for
the $I_{r1}$ involve only the products $\pi_i \pi_i^*$, which
correspond to the Pl\"ucker coordinates for the zero weight space in
$P({\bf C}^n \times \wedge^{n-1} {\bf C}^n)$, whereas in Case B, the $I_{r1}$
depend only on the Pl\"ucker coordinates $\pi_i \pi_j^*$ with $i > j$,
corresponding to the negative roots of $sl(n,{\bf C})$.
We point out, however, that we are using different embeddings in the
two cases.  Later we show that using the companion embedding for a
level set with distinct eigenvalues, none of which is zero, the
expressions for the 1-chop 
integrals involve all the weights of $sl(n,{\bf C})$; this,
however, obscures the geometry of the base locus defined below, which
is revealed clearly by the torus embedding.

\begin{defn} Let ${\cal F}^0_{r, \alpha}$ for $\alpha \in {\bf C}$ denote
the variety in 
$G/P_1$ cut out by the homogeneous polynomial
\begin{equation}
I_{r1}(\sum_{i=1}^{n-1} \pi_{i+1} \pi_{i}^*) - \sum_{i=1}^{n-r-1}
\pi_{i+r+1} \pi_i^* = 0, \label{F0}
\end{equation}
where $I_{r1} = \alpha$, and let ${\cal F}^0_{r, \infty}$ denote the
variety
\begin{equation}
\sum_{i=1}^{n-1} \pi_{i+1} \pi_{i}^* = 0.
\end{equation}
The base locus of these varieties is
the intersection
$${\cal Z}_0 = \bigcap_{r=1}^{n-2} \bigcap_{\alpha \in \bf{P^1}}  
{\cal F}^0_{r, \alpha}.$$
\end{defn}

This base locus corresponds to the subset of $(\epsilon + {\cal
B}_-)_0$ on which all the $I_{r1}$ are undefined (and not equal to
infinity).  It consists of the $n-1$ components 
\begin{equation}
\pi_1^* = \cdots = \pi_k^* = \pi_{k+1} = \cdots = \pi_n = 0
\label{Z0kcomp}
\end{equation}
for $k = 1, \ldots, n-1$; we denote them by 
${\cal Z}_0^k$.  Each of these components is the
closure of a single torus orbit in $G/P_1$ of complex dimension $n-2$.
To see this, observe that any two generic points in  
${\cal Z}_0^k$ have Pl\"ucker coordinates of the form 
$[\pi_1 : \cdots : \pi_k : 0 : \cdots : 0] \times [0 : \cdots : 0 :
\pi_{k+1}^* : \cdots : \pi_n^*]$ and 
$[\pi'_1 : \cdots : \pi'_k : 0 : \cdots : 0] \times [0 : \cdots : 0 :
\pi_{k+1}^{*'} : \cdots : \pi_n^{*'}]$, where $\pi_1 \cdots \pi_k
\pi_{k+1}^* \cdots \pi_n^* \neq 0$ and $\pi'_1 \cdots \pi'_k
\pi_{k+1}^{*'} \cdots \pi_n^{*'} \neq 0$.  Since the element in the torus
whose diagonal entries are $h_i = {\pi'_i \over \pi_i}$ for $i = 1,
\ldots,k$ and $h_i = {\pi_i^* \over \pi_i^{*'}}$ for $i = k+1, \ldots,
n$ takes the first into the second, these points belong to the same torus
orbit; its dimension is $n-2$.

The geometry of ${\cal Z}_0$ has a simple description in
terms of the momentum mapping on $G/P_1$ and the momentum polytope
$\triangle_n$.  

\begin{defn}
Let $Q$ be the weight polytope of an irreducible representation of
$Sl(n,{\bf C})$ with highest weight $w$.  We denote by $Star(Q,w)$ the union
of all faces 
which have $w$ as a vertex and are perpendicular to a fundamental
weight of $sl(n,{\bf C})$ with respect to the Killing form on ${\cal H}^*$.
We denote the fundamental weight $\sum_{i=1}^k L_i$ by $w_k$ for $k = 1,
\ldots, n-1$.
\end{defn}

\noindent {\bf Remarks:}  1.  The face of $Star(Q,w)$ perpendicular to
$w_k$ intersects the positive ray through $w_k$ perpendicularly at its
center.  

2.  For an arbitrary irreducible representation, $Star(Q,w)$ is the
union of at most $n-1$ faces of dimension $n-2$.  In the case of
$Star(\triangle_n, L_1 - L_n)$, this number is equal to $n-1$.

\begin{defn} \label{subalgebras} Let $S_k$ denote the set $\{1,
\cdots, k\}$, and for $R 
\subset S_n$ with $|R| = k$, denote by $R'$ the complement $S_n
\setminus R$.  For each $R \subset S_n$, we define ${\cal G}_R$ to be the
subalgebra of $sl(n,{\bf C})$ generated by the roots $L_i - L_j$ for $i,j
\in R$.  (${\cal G}_R \cong sl({|R|},{\bf C})$.)  The subalgebras
${\cal G}_R \oplus 
{\cal G}_{R'}$ for $R \subset S_n$ with $1 \leq |R| \leq n-1$ have rank
$n-2$.  We will refer to these as the rank $n-2$ principal subalgebras
of $sl(n,{\bf C})$.  For $R = S_k$, $k = 1, \ldots, n-1$, the subalgebras
${\cal G}_{S_k} \oplus {\cal G}_{S'_k} \cong sl(k,{\bf C}) \oplus
sl(n-k,{\bf C})$ will be 
called the rank $n-2$ principal block subalgebras of $sl(n,{\bf C})$.
\end{defn}

\begin{prop} \label{BIr1} Let $\mu$ be the momentum mapping on $G/P_1$.  Then 
${\cal Z}_0 = \mu^{-1}(Star(\triangle_n, L_1 - L_n))$.  The
momentum mapping induces a one-to-one correspondence between the $n-1$
components of ${\cal Z}_0$, each of which is the closure of a
single torus orbit of complex dimension $n-2$, and the set of
$(n-2)$-dimensional faces of $Star(\triangle_n, L_1 - L_n)$; $\mu({\cal
Z}_0^k)$ is the face of $Star(\triangle_n, L_1 - L_n)$
perpendicular to the fundamental weight $w_k$.
\end{prop}

\noindent {\bf Proof.}  We have already shown that ${\cal
Z}_0^k$ is the closure of a single torus orbit in $G/P_1$ of
complex dimension $n-2$.  Its image under the momentum mapping is
therefore an $(n-2)$-dimensional face of $\triangle_n$ whose vertices
are the images of the fixed points of the torus action contained in
it.  These fixed points are those whose single nonvanishing product of
the form $\pi_i \pi_j^*$ is such that $i \in S_k$ and $j \in S'_k$;
their images under $\mu$ are the (positive) roots $L_i - L_j$ for
$i \in S_k$ and $j \in S'_k$.  These are precisely the weights of the
representation $V_{S_k} \otimes V_{S'_k}^*$ of the subalgebra 
${\cal G}_{S_k} \oplus {\cal G}_{S'_k}$, where $V_{S_k}$ and
$V_{S'_k}^*$ denote the 
standard representation of ${\cal G}_{S_k}$ and the dual of the standard
representation of ${\cal G}_{S'_k}$, respectively.  This representation is
the irreducible component containing $e_1 \otimes e_n^*$ in the
decomposition   
of the adjoint representation of $sl(n,{\bf C})$ under the action of this
subalgebra.  Since the root lattice of ${\cal G}_{S_k} \oplus {\cal
G}_{S'_k}$ is 
generated by the simple roots $L_i - L_{i+1}$ for $i \neq k$, which
are all perpendicular to the fundamental weight $w_k$ with respect to
the Killing form, $\mu({\cal Z}_0^k)$ is the face of 
$Star(\triangle_n, L_1 - L_n)$ perpendicular to $w_k$.
\rule[-1mm]{1mm}{2mm}

\bigskip

\noindent {\bf Remark:} ${\cal Z}_0^k$, being the closure of
a single torus orbit in $G/P_1$, is a toric variety.  The geometry of
such varieties in homogeneous spaces $G/P$ has been studied in detail
in \cite{FH}; see also the corrections in \cite{Dabrowski}.  The
polytope $\mu({\cal 
Z}_{0}^k)$, which is the weight polytope of the representation 
 $V_{S_k} \otimes V_{S'_k}^*$ of ${\cal G}_{S_k} \oplus {\cal
G}_{S'_k}$, contains  
significant information about the geometry of this variety in $G/P_1$.

\bigskip

For Case A, we have a result similar to Proposition~\ref{BIr1}.

\begin{defn} Let ${\cal F}^{\Lambda}_{r, \alpha}$ for $\alpha \in \bf{C}$
denote the 
variety in $G/P_1$ cut out by the homogeneous polynomial
\begin{equation} 
I_{r1}(\sum_{i = 1}^n \sigma_1(\hat i) \pi_i \pi_i^*) -  (\sum_{i=1}^n
\sigma_{r+1}(\hat i) \pi_i \pi_i^*) = 0,
\label{Flambda}  \end{equation}
where $I_{r1} = \alpha$, and let ${\cal F}^{\Lambda}_{r, \infty}$
denote the variety
\begin{equation}
\sum_{i = 1}^n \sigma_1(\hat i) \pi_i \pi_i^* = 0. \label{Flambdainf}
\end{equation} 
The base locus of these varieties is
the intersection
$${\cal Z}_{\Lambda} = \bigcap_{r=1}^{n-2} \bigcap_{\alpha \in
\bf{P^1}} {\cal F}^{\Lambda}_{r, \alpha}.$$
\end{defn}

${\cal Z}_{\Lambda}$ corresponds (under the
torus embedding) to the subset of $(\epsilon + {\cal B}_-)_{\Lambda}$
on which all the 1-chop integrals are undefined (and not equal to
infinity).  Since the common vanishing set of equations
(\ref{Flambda}) and (\ref{Flambdainf})
coincides with the intersections of the varieties $\pi_i \pi_i^* = 0$
for $i = 1, \ldots, n$, ${\cal Z}_{ \Lambda}$  is the union of the
$\sum_{k=1}^{n-1} {n \choose k}$ components
\begin{equation} \pi_{\sigma(1)}^* = \cdots = \pi_{\sigma(k)}^* = 
\pi_{\sigma(k+1)} = \cdots = \pi_{\sigma(n)} = 0 \label{Acomponent}
\end{equation}
for $k = 1, \ldots, n-1$ and $\sigma \in \Sigma_n$, the Weyl group of
$sl(n,{\bf C})$.  We denote the component (\ref{Acomponent}) by ${\cal
Z}_{\Lambda}^{k, \sigma}$.  (Note that the permutation
$\sigma$ is not unique for $n \geq 3$.)

In this case, as in Case B, one can easily show that each of these
components is the closure of a single torus orbit in $G/P_1$ of
complex dimension $n-2$.  Its image under the momentum mapping is
therefore an $(n-2)$-dimensional face of $\triangle_n$.  In fact, all
such faces in the boundary of $\triangle_n$ are obtained in this way.

\begin{prop} \label{AIr1}
${\cal Z}_{\Lambda} = \mu^{-1}(\partial \triangle_n)$.  The
momentum mapping induces a one-to-one correspondence between the
components of ${\cal Z}_{\Lambda}$, each of which is the
closure of a single torus orbit of complex dimension $n-2$, and the set
of all $(n-2)$-dimensional faces of $\triangle_n$; $\mu({\cal
Z}_{\Lambda}^{k, \sigma})$ is the face of $\triangle_n$ which
intersects the positive ray through the weight $\sigma(w_k) =
\sum_{i=1}^k L_{\sigma(i)}$ perpendicularly.
\end{prop}

\noindent {\bf Proof.} By Proposition~\ref{BIr1}, $\mu({\cal
Z}_{0}^k)$ is the face of $\triangle_n$ which intersects the
positive ray through $w_k$ perpendicularly at its center.  Its
vertices are $L_i - L_j$ for $i \in S_k, j \in S'_k$.  The action of
$\sigma \in \Sigma_n$ on the roots of $sl(n,C)$ takes this face to the
face of $\triangle_n$ whose vertices are $L_i - L_j$ for $i \in
\sigma(S_k)$ and $j \in \sigma(S'_k)$; it intersects the positive
ray through the weight $\sigma(w_k) = \sum_{i=1}^k L_{\sigma(i)}$
perpendicularly.  But these vertices are the images of the fixed
points of the torus action in $G/P_1$ for which the nonvanishing
product $\pi_i \pi_j^*$ is such that $i \in \sigma(S_k)$ and $j \in
\sigma(S'_k)$, which are precisely the fixed points
contained in the component ${\cal Z}_{\Lambda}^{k, \sigma}$ of
${\cal Z}_{\Lambda}$. The image $\mu({\cal Z}_{\Lambda}^{k, 
\sigma})$  is therefore the $(n-2)$-dimensional face of $\triangle_n$
which intersects the positive ray through $\sigma(w_k)$
perpendicularly. 

To complete the proof, it remains to show that every
$(n-2)$-dimensional face of $\triangle_n$ is perpendicular to a weight
of a fundamental representation of $sl(n,{\bf C})$.  To see this, observe
that the vertices of the face which intersects the positive ray
through the weight 
$\sigma(w_k)$ perpendicularly are the weights of the representation
$V_{\sigma(S_k)} 
\otimes V_{\sigma(S'_k)}^*$ of the principal rank $n-2$ subalgebra
${\cal G}_{\sigma(S_k)} \oplus {\cal G}_{\sigma(S'_k)}$ of
$sl(n,{\bf C})$, where 
$V_{\sigma(S_k)}$ and $V_{\sigma(S'_k)}^*$ are the standard
representation of ${\cal G}_{\sigma(S_k)}$ and the dual of the standard
represention of ${\cal G}_{\sigma(S'_k)}$, respectively.  If we denote
the weight 
polytopes of $V_{\sigma(S_k)}$ and $V_{\sigma(S'_k)}^*$ by
$Q_{\sigma(S_k)}$ and  $Q_{\sigma(S'_k)}^*$, respectively, then this
face is the polytope $Q_{\sigma(S_k)} \times Q_{\sigma(S'_k)}^*$, the
Cartesian product of a $k$-simplex and an $(n-k)$-simplex.  One can
easily write down an arbitrary $(n-3)$-dimensional face of this
polytope and verify that it is also a face of another such
$(n-2)$-dimensional face of $\triangle_n$.
\rule[-1mm]{1mm}{2mm}

\bigskip

\noindent {\bf Remark:} In Case A, the image $\mu({\cal Z}_{
\Lambda})$ of the base locus ${\cal Z}_{\Lambda}$ contains all
the roots of $sl(n,{\bf C})$, whereas in Case B, $\mu({\cal Z}_0)$
contains precisely the vertices of $\triangle_n$ corresponding to the
positive roots.  Compare this with the remark following
Proposition~\ref{BIr1expr}.

\subsection{The companion embedding for an arbitrary isospectral level
set} 

Let $C$ be the companion matrix defined in
Proposition~\ref{normalform} for an arbitrary choice of spectrum.
From the proof of 
Propsition~\ref{BIr1expr}, 
\begin{equation}
det(L(C - \lambda I)L^{-1})_{(1)} = \sum_{i,j = 1}^n (-1)^{i+j}
\pi_i  \pi_j^* (C - \lambda I)_{\hat i \hat j}. \label{generalcomp}
\end{equation}
Recall that in Case B, in which all the eigenvalues are zero, all the
symmetric functions vanish, and the
coefficients in this polynomial involve 
only the Pl\"ucker coordinates corresponding to the negative roots of
$sl(n,{\bf C})$.  However, on an isospectral level set for which not all
eigenvalues are zero, the symmetric functions $s_2, \ldots, s_n$ 
do not all vanish, and the coefficients of the polynomial
(\ref{generalcomp}) involve additional terms which depend on the
values of these functions.  Rewriting (\ref{generalcomp}) by
collecting the coefficients of the $s_i$, one obtains

\vspace{4cm}

\begin{eqnarray*}
(-1)^{n+1}det(L(C - \lambda I)L^{-1})_{(1)}& = &\sum_{i=1}^{n-1}
\big( \lambda^{n-1-i} \sum_{ht(\alpha) = -i} \pi_{\alpha} \big) \\
& + &\sum_{p=2}^n s_p \Big[ \sum_{i=1}^{p-1} \big( \lambda^{n-1-i}
\sum_{\begin{array}{c} ht(\alpha) = p-i \\  \alpha \in \Gamma({\cal
G}_{S'_{n-p}}) \end{array}} \pi_{\alpha} \big) \\
&  & - \lambda^{n-1-p} \sum_{i=1}^{n-p} \pi_i \pi_i^* \\
&  & - \sum_{i=1}^{n-1-p}
\big( \lambda^{n-1-p-i} \sum_{\begin{array}{c} ht(\alpha) = -i \\
\alpha \in \Gamma({\cal G}_{S_{n-p}}) \end{array}} \pi_{\alpha}
\big) \Big],  
\label{generalcomp2}
\end{eqnarray*}
where $\Gamma({\cal G}_R)$ denotes the sublattice of the root lattice
of $sl(n,{\bf C})$ determined by the subalgebra ${\cal G}_R$ (see
definition~\ref{subalgebras}). This expression is calculated easily in
low-dimensional examples and is verified in general by induction on
$n$.  It is the sum of $n$ polynomials of degree $n-2$; the first is
the polynomial $det(L(C_0 - \lambda I)L^{-1})_{(1)}$ of Case B, and each
of the others is multiplied by a symmetric polynomial $s_i$.  

Observe that the coefficients of the polynomial containing the factor
$s_p$ correspond to the subalgebra ${\cal G}_{S_{n-p}} \oplus 
{\cal G}_{S'_{n-p}}$ of $sl(n,{\bf C})$.  The coefficients of
$\lambda^{n-2}, \ldots, \lambda^{n-p}$ are the sums of the Pl\"ucker
coordinates for the (positive) roots of heights $p-1,
\ldots, 1$, respectively, of the rank $p-1$ subalgebra 
${\cal G}_{S'_{n-p}}$; the coefficient of $\lambda^{n-p-1}$ is (minus)
the sum of the Pl\"ucker coordinates for the zero weight space of
${\bf C}^{n-p} \otimes ({\bf C}^{n-p})^*$, and the coefficients of 
$\lambda^{n-p-2}, \ldots, \lambda^{0}$ are (minus) the sums of the Pl\"ucker
coordinates for the (negative) roots of heights $-1,
\ldots, -(n-p-1)$, respectively, of the rank $n-p-1$ subalgebra  
${\cal G}_{S_{n-p}}$.

For purposes of illustration, consider the case $n=4$:
\begin{eqnarray*}
-det(L(C - \lambda I)L^{-1})_{(1)}& = &\lambda^2(\pi_2 \pi_1^* + 
\pi_3 \pi_2^* + \pi_4 \pi_3^*) + \lambda(\pi_3 \pi_1^* + \pi_4
\pi_2^*) + \pi_4 \pi_1^* \\
& + &s_2[\lambda^2 \pi_3 \pi_4^* - \lambda(\pi_1 \pi_1^* + \pi_2
\pi_2^*) - \pi_2 \pi_1^*] \\
& + &s_3[\lambda^2 \pi_2 \pi_4^* + \lambda(\pi_2 \pi_3^* + \pi_3
\pi_4^*) - \pi_1 \pi_1^*] \\
& + &s_4[\lambda^2 \pi_1 \pi_4^* + \lambda(\pi_1 \pi_3^* + \pi_2
\pi_4^*) + (\pi_1 \pi_2^* + \pi_2 \pi_3^* + \pi_3 \pi_4^*)].
\end{eqnarray*}
When $s_2, s_3$, and $s_4$ all vanish, this reduces to the polynomial 
of Case B, whose coefficients
depend only on the Pl\"ucker coordinates for the negative
roots of $sl(4,{\bf C})$.  The
coefficients of the polynomial containing the factor $s_2$ correspond
to the subalgebra ${\cal G}_{\{1,2\}} \oplus {\cal G}_{\{3,4\}}$.
The constant term is (minus) the Pl\"ucker coordinate for the negative
root of ${\cal G}_{\{1,2\}}$ (of height -1), the coefficient of $\lambda$
is (minus) the sum of the  Pl\"ucker coordinates for the 
zero weight space of ${\bf C}^2~\otimes~({\bf C}^2)^*$, and the coefficient of
$\lambda^2$ is the  Pl\"ucker coordinate for the positive root of
${\cal G}_{\{3,4\}}$ (of height 1).  This pattern repeats; the
coefficients of the polynomial containing the factor $s_3$ correspond
to the subalgebra ${\cal G}_{\{1\}} \oplus {\cal G}_{\{2,3,4\}}$ in the same
way.  The constant term is (minus) the Pl\"ucker coordinate for the
zero weight space of ${\bf C} \otimes {\bf C}^*$, and the coefficients of
$\lambda$ and $\lambda^2$ are the sum of the roots of ${\cal
G}_{\{2,3,4\}}$ of height 1 and the root of ${\cal G}_{\{2,3,4\}}$ of height
2, respectively.  Finally, the coefficients of $\lambda^2, \lambda$,
and $\lambda^0$ in the polynomial
corresponding to $s_4$ are simply the sums of the Pl\"ucker
coordinates for the positive roots of $sl(4,{\bf C})$ ($= {\cal
G}_{\{1,2,3,4\}}$) of heights 3, 2, and 1, respectively.

Writing the polynomial $det(L(C  - \lambda I)L^{-1})_{(1)}$ in this
way allows one to see how the expressions of the 1-chop integrals in
the case of the companion embedding simplify as a generic isospectral
level set approaches the extreme case in which all eigenvalues are
zero through a sequence of degenerations in which one additional eigenvalue
is set equal to zero in each step.  Indeed, the $m+1$ symmetric functions
$s_{n-m}, \ldots, s_n$ vanish simultaneously if and only if zero is an
eigenvalue of multiplicity at least $m+1$.  In the case that all the
$\lambda_i$ are nonzero, $s_n \neq 0$, and the coefficients of the
1-chop polynomial depend on the Pl\"ucker coordinates for all the roots
of $sl(n,{\bf C})$.  With each additional eigenvalue that is set equal to
zero, the 1-chop polynomial loses its dependence on the Pl\"ucker
coordinates for the roots of the greatest height that
occurs in the previous step so that when $m$ eigenvalues vanish, the
1-chop integrals do not depend on the roots of height greater than or
equal to $n-m$.

\section{The Special Case $n=3$}

In this example, we take $G = Sl(3,{\bf C})$.  The elements of $\epsilon +
{\cal B}_-$ have the form 
$$X = \left( \begin{array}{ccc} f_1 & 1 & 0 \\ g_1 & f_2 & 1 \\ h &
g_2 & f_3 \end{array} \right), \qquad \sum_{i = 1}^3 f_i = 0,$$  
and there is a single 1-chop integral, ${\cal C} = f_2 - {g_1g_2
\over h}$, which is a Casimir.   In this case $G/P_1$ coincides with
the flag manifold $G/B$, and the momentum polytope is the regular
hexagon shown in Figure 1.
  
In Case A, the Casimir has the expression 
\begin{equation}
{\cal C} = \frac{\lambda_2 \lambda_3 \pi_1 \pi_1^* +  \lambda_1
\lambda_3 \pi_2 \pi_2^* + \lambda_1 \lambda_2 \pi_3 \pi_3^*}{\lambda_1
\pi_1 \pi_1^* + \lambda_2 \pi_2 \pi_2^* + \lambda_3 \pi_3 \pi_3^*}.
\label{CasimirA} \end{equation}
The base locus  ${\cal Z}_{\Lambda}$ is the common intersection
of the varieties ${\cal F}_{\cal C}^{\Lambda}$ defined by
$${\cal C} (\sum_{i=1}^3 \sigma_1(\hat i) \pi_i \pi_i^*) - \sum_{i=1}^3
\sigma_2(\hat i) \pi_i \pi_i^* = 0$$
for ${\cal C} \in {\bf C}$ and by the vanishing of the denominator in
(\ref{CasimirA}) for ${\cal C} = \infty$.
It is the union of the strata in the flag
manifold whose images under the momentum mapping lie on the boundary
of the momentum polytope.  Each of these strata consists of a unique
torus orbit; there are six one-dimensional complex orbits
corresponding to the 
edges of the hexagon and six fixed points whose images are the vertices.
 
This base locus and its image
under the momentum mapping precisely encode the structure of the
intersection of the isospectral submanifold $(\epsilon + {\cal
B}_-)_{\Lambda}$ with the
set of symplectic leaves in $\epsilon + {\cal B}_-$ of complex
dimension strictly less than four, the dimension of the generic
leaves. Each of the six 1-dimensional orbits in ${\cal
B}_{\Lambda}$ corresponds under the torus embedding 
to the intersection of $(\epsilon + {\cal B}_-)_{\Lambda}$ with a
symplectic leaf of complex dimension two; the six fixed points of the
torus action are the images of the six 0-dimensional leaves contained in 
$(\epsilon + {\cal B}_-)_{\Lambda}$.  In terms of the 
momentum polytope, the intersections of the lower-dimensional leaves
with $(\epsilon + {\cal B}_-)_{\Lambda}$ correspond to the
six edges and the six vertices of the hexagon.
This is explained in detail and illustrated in Chapter
Three of \cite{diss}. 

The expression for the Casimir in Case B is
\begin{equation}
{\cal C} = \frac{\pi_3 \pi_1^*}{\pi_2 \pi_1^* + \pi_3 \pi_2^*}.
\label{CasimirB} \end{equation}
The base locus ${\cal Z}_0$ is the union of
the closures of two 1-dimensional complex torus orbits in $G/B$, one
satisfying 
$\pi_1^* = \pi_2 = \pi_3 = 0$ and the other satisfying $\pi_1^* =
\pi_2^* = \pi_3 = 0$.  Their
images under the momentum mapping are the two edges of
$Star(\triangle_3, L_1 - L_3)$, shown in Figure 2.

As in Case A, the intersection of the isospectral submanifold 
$(\epsilon + {\cal B}_-)_0$ with the set of 
symplectic leaves of dimension less than four corresponds to
the base locus ${\cal Z}_0$.  $(\epsilon + {\cal B}_-)_0$   
has nontrivial intersections with only two of the
two-dimensional leaves; these intersections have the forms
$$\left( \begin{array}{ccc} a & 1 & 0 \\ b & -a & 1 \\ 0 & 0 & 0
\end{array} \right) \qquad \mbox{and} \qquad \left( \begin{array}{ccc}
0 & 1 & 0 \\ 0 & a & 1 \\ 0 & b & -a \end{array} \right)$$
with $a^2 -b = 0$, which  correspond under the companion embedding to
the two 1-dimensional complex torus orbits in ${\cal Z}_0$. The 
unique 0-dimensional leaf in $(\epsilon + {\cal B}_-)_0$ is the
companion matrix.  Its image under the companion embedding is the
fixed point of the torus action whose image under the moment mapping
is the highest weight $L_1 - L_3$, in which the two edges of
$Star(\triangle_3, L_1 - L_3)$ intersect. 

Consider now the variety ${\cal F}_{\cal C}^0$ defined by
$${\cal C} (\pi_2 \pi_1^* + \pi_3 \pi_2^*) - \pi_3
\pi_1^* = 0$$
for ${\cal C} \in {\bf C}$ and by the vanishing of the denominator in
({\ref{CasimirB}) for ${\cal C} = \infty$.  For ${\cal C} = 0$, ${\cal
F}^0_0$ is the union of the two components $\pi_3 = 0$ and $\pi_1^* =
0$.  It is easy to show that each of these components is the closure
of a single nongeneric two-dimensional complex torus orbit in $G/P_1$.
Their images under the momentum mapping are the two half-hexagons
shown in Figure 3.  For a nonzero value of the Casimir, the variety
${\cal F}_{\cal C}^0$ is not invariant under the torus action.  The
image of ${\cal F}_{\cal C}^0$ under the momentum mapping for ${\cal
C} \not \in \{0, \infty\}$ contains exactly two 
points on the boundary of the momentum polytope which are not in the
base locus.  These are the points ${1 \over 1 + |{\cal
C}|^2}~[~(L_2~-~L_1) + |{\cal C}|^2(L_3 - L_1)]$ and ${1 \over 1 +
|{\cal 
C}|^2} [(L_3 - 
L_2) + |{\cal C}|^2(L_3 - L_1)]$, which lie on the two edges opposite
the edges of $Star(\triangle_3, L_1 - L_3)$.  They are the images of
$[0 : 1 : {\cal C}] \times [1 : 0 : 0]$ and $[0 : 0 : 1] \times [{\cal C}
: 1 : 0]$, respectively, in ${\cal F}_{\cal C}^0$.  Observe that each
of these points lies at the same distance from the common vertex $L_3
- L_1$ of these two edges and that this distance depends only on the
modulus of ${\cal C}$.  As $|{\cal C}|^2$ increases from $0$ to $\infty$,
the two points move monotonically along the two edges, starting at the
simple negative roots $L_2 - L_1$ and $L_3 - L_2$ when ${\cal C} = 0$
and coinciding in the limit at $L_3 - L_1$ when ${\cal C} = \infty$.

Furthermore, each of the two edges opposite $Star(\triangle_3, L_1 -
L_3)$ is the image under $\mu$ of the closure of a unique
1-dimensional complex torus orbit in $G/P_1$, which is a copy of ${\bf P}^1$.
These two copies intersect in the fixed point for which $\pi_3 \pi_1^*
\neq 0$; they are parametrized as $[0
: z : w] \times [1 : 0 : 0]$ and $[0 : 0 : 1] \times [z : w : 0]$ for
$[z : w] \in {\bf P}^1$.  Since each of these copies of the projective line
intersects ${\cal F}_{\cal C}^0$ in a unique point for every ${\cal C}
\in {\bf P}^1$, it can be considered as the parameter space for
the varieties ${\cal F}_{\cal C}^0$; the two fixed points of the torus
action contained in it correspond to ${\cal C} = 0$ and ${\cal C} =
\infty$. 
Thus, the modulus of ${\cal C}$ determines the image of each point in
this parameter space under the momentum mapping so that each orbit of
the compact torus $S^1$ in ${\bf P}^1$ has the same image under $\mu$.

\bigskip

\noindent {\bf Remark:} The geometry of the base locus ${\cal
Z}_0$ and the reducible variety ${\cal F}_{\cal C}^0$ 
is seen in the Bruhat decomposition of the flag manifold,
$$ G/B = \bigcup_{w \in \Sigma_3} BX_wB/B,$$ where $X_w$ is a matrix
representing the permutation $w$.  The single zero-dimensional cell is
the fixed point in  ${\cal Z}_0$ whose image under $\mu$ is the highest
weight 
$L_1 - L_3$, corresponding to the unique zero-dimensional leaf in 
$(\epsilon + {\cal B}_-)_0$. The two one-dimensional complex cells are
the torus 
orbits in ${\cal Z}_0$ corresponding to the two edges of  
$Star(\triangle_3, L_1 - L_3)$. The two two-dimensional
complex cells are the components of the reducible variety ${\cal
F}_0^0$, and the big cell contains the union of the generic level
set varieties ${\cal F}_{\cal C}^0$ (minus the base locus) for ${\cal
C}\in {\bf P}^1 \setminus \{0\}$.

For a geometrical study of the varieties 
${\cal F}_{\cal C}^{\Lambda}$ in Case A, we refer to Chapter Three of
\cite{diss}, in which it is shown that the geometry of the base locus
${\cal Z}_{\Lambda}$, the three reducible varieties
${\cal F}_{\lambda_i}^{\Lambda}$, and the irreducible varieties
${\cal F}_{\cal C}^{\Lambda}$ for ${\cal C} \neq \lambda_i$, is precisely
encoded in the partitioning of 
the flag manifold into strata (see \cite{GelfandSerganova}).  This
stratification of $G/B$ is finer than the Bruhat decomposition which,
as we have shown,
describes the analogous geometry in Case B.

%\begin{figure}[p]
%\vspace{18cm}
%\end{figure}

\begin{figure}[t]
\vspace{11cm} 
\end{figure}

\section{The Special Case $n=4$}

In the full $Sl(4,{\bf C})$ Kostant-Toda lattice, there are two 1-chop
integrals, $I_{11}$, which is a Casimir, and the constant of motion
$I_{21}$.  In Case A, these integrals have the expressions
$$I_{11} = \frac{\sum_{i=1}^4 \sigma_2(\hat i) \pi_i \pi_i^*}
{\lambda_1 \pi_1 \pi_1^* + \lambda_2 \pi_2 \pi_2^* + \lambda_3 \pi_3
\pi_3^* + \lambda_4 \pi_4 \pi_4^*},$$
$$I_{21} = - \, \frac{\lambda_2 \lambda_3 \lambda_4 \pi_1 \pi_1^* + 
\lambda_1 \lambda_3 \lambda_4 \pi_2 \pi_2^* + 
\lambda_1 \lambda_2 \lambda_4 \pi_3 \pi_3^* +
\lambda_1 \lambda_2 \lambda_3 \pi_4 \pi_4^*}{\lambda_1 \pi_1 \pi_1^* +
\lambda_2 \pi_2 \pi_2^* + \lambda_3 \pi_3 
\pi_3^* + \lambda_4 \pi_4 \pi_4^*}.$$
The momentum polytope $\triangle_4$ of $G/P_1$ is the cuboctahedron,
shown in Figure 4; it has fourteen two-dimensional faces.  The base
locus ${\cal Z}_{\Lambda}$ is the union of fourteen components, each of
which is the closure of the unique two-dimensional complex torus orbit
in $G/P_1$ whose image under $\mu$ is a particular two-dimensional
face of the momentum polytope.  

In Case B  the 1-chop integrals are expressed as
$$I_{11} = \frac{\pi_3 \pi_1^* + \pi_4 \pi_2^*}{\pi_2 \pi_1^* + \pi_3
\pi_2^* + \pi_4 \pi_3^*},$$
$$I_{21} = \frac{\pi_4 \pi_1^*}{\pi_2 \pi_1^* + \pi_3 \pi_2^* + \pi_4
\pi_3^*}.$$  The base locus, ${\cal Z}_0$, of the corresponding
varieties in $G/P_1$ is the union of the closures
of three two-dimensional complex torus orbits. Its
image under the momentum mapping is $Star(\triangle_4, L_1 - L_4)$,
which consists of the two triangular faces perpendicular to the
fundamental weights $L_1$ 
and $-L_4$ and the square face perpendicular to the fundamental weight
$L_1 + L_2$, as illustrated in Figure 5.

The isomorphism $\rho : sl(4,{\bf C}) \rightarrow so(6,{\bf C})$ gives rise to
another constant of motion, $J$, which is in involution with the
integrals ${1 \over k}trX^k, k = 2,3,4$, but not with $I_{21}$.  It is
obtained by a chopping construction on the matrix $\rho(X) - \lambda
I$ similar to the one on $X - \lambda I$ described in
Proposition~\ref{k-chopcalc} 
which gives the 1-chop integrals $I_{r1}$.  The restriction of $J$ to
the generic symplectic leaf on which the Casimir $I_{11}$ is equal to
zero exhibits a geometry very similar to that of the level sets of the
1-chop integrals in each of the two types of isospectral submanifolds
we have been considering.

We start with Case A.  Let ${\cal V}_{\Lambda}$ be the subvariety
of the flag manifold defined by
$$\sum_{i=1}^4 \sigma_2(\hat i) \pi_i \pi_i^* = 0,$$ on which 
$I_{11}$ vanishes.  In \cite{diss} it is found that the
function $J$, when restricted to ${\cal V}_{\Lambda}$, depends only on its
projection to the Grassmannian $G(2,4)$ of two-dimensional subspaces
of ${\bf C}^4$.  $G(2,4)$ is the orbit $Sl(4,{\bf
C})~\cdot~[e_1~\wedge~e_2] 
\subset 
P(\wedge^2 {\bf C}^4)$.  The weights of the irreducible representation
$\wedge^2 {\bf C}^4$ of $sl(4,{\bf C})$ are $L_i + L_j$ for $i \neq j$.  We 
will take the Pl\"ucker coordinates on $G(2,4)$ with respect to the
weight basis 
$\{e_i \wedge e_j\}_{i < j}$ and denote $\pi_{L_i + L_j}$ by $\pi_{ij}$.
In terms of these coordinates, the expression for the restriction of
$J$ to ${\cal V}_{\Lambda}$ as given in \cite{diss} is
$$J|_{{\cal V}_{\Lambda}} = 
\frac{A_{\Lambda} A'_{\Lambda} \pi_{12} \pi_{34} - B_{\Lambda} B'_{\Lambda}
\pi_{13} \pi_{24}}{A_{\Lambda} \pi_{12} \pi_{34} - B_{\Lambda} \pi_{13}
\pi_{24}},$$
where
\begin{eqnarray*} A_{\Lambda} & = & (\lambda_1 - \lambda_3)(\lambda_2 -
\lambda_4), \\
A'_{\Lambda} & = & (\lambda_1 + \lambda_3)(\lambda_2 + \lambda_4), \\
B_{\Lambda} & = & (\lambda_1 - \lambda_2)(\lambda_3 - \lambda_4), \\
B'_{\Lambda} & = & (\lambda_1 + \lambda_2)(\lambda_3 + \lambda_4).
\end{eqnarray*}
This is simply a linear fractional transformation of the cross-ratio
$c$ defined by
$$c = \frac{\pi_{12} \pi_{34}}{\pi_{13} \pi_{24}}.$$

Let $\tilde{\cal Z}_{\Lambda}$ denote the common intersection of the
varieties $$c \, \pi_{13} \pi_{24} - \pi_{12} \pi_{34} = 0$$ in $G(2,4)$
for $c \in {\bf C}$, and let $\mu_{G(2,4)}$ be the momentum mapping on the
Grassmannian.  The momentum polytope, which we denote by
$\Box$, is 
shown in Figure 6.  Our statement in this situation is parallel to
Proposition~\ref{AIr1} for the integrals $I_{r1}$ in Case A.  Its proof is
found in \cite{diss}.  

\begin{prop} $\tilde{\cal Z}_{\Lambda} =  \mu_{G(2,4)}^{-1}(\partial
\Box)$.  The momentum mapping induces a one-to-one correspondence
between the eight components of $\tilde{\cal Z}_{\Lambda}$, each of which
is the closure of a single torus orbit of complex dimension two, and
the set of two-dimensional faces of $\Box$.  
\end{prop}

Consider now Case B, in which we take the companion embedding of 
$(\epsilon + {\cal B}_-)_0$ into $G/B$, and let ${\cal V}_0$ be the
subvariety of $G/B$ defined by
$$\pi_3 \pi_1^* + \pi_4 \pi_2^* = 0,$$
on which $I_{11}$ vanishes.  The expression for $J$ restricted to
${\cal V}_0$ is
\begin{equation}
J|_{{\cal V}_0} = \frac{\pi_{34}^2}{\pi_{34}(\pi_{14} + \pi_{23}) -
\pi_{24}^2}. \label{J0}
\end{equation}
(To see this, it suffices to show that it holds on the matrices
$X(I_{21},u)$ defined by equation (5.9) in \cite{diss} with all
eigenvalues set equal to zero since $J$ is invariant under the torus
action and $u$ parametrizes the generic torus orbits in each generic
level set of $I_{21}$ in $(\epsilon + {\cal B}_-)_0$.  Indeed,
evaluating our formula (\ref{J0}) on the image under the companion
embedding of $X(I_{21},u)$ with the eigenvalues equal to zero
gives precisely the expression for $J$ in equation (6.4) of
\cite{diss} with all eigenvalues zero.)

Let $\tilde{\cal Z}_{0}$ denote the common intersection of the varieties
$$J|_{{\cal V}_0}(\pi_{34}(\pi_{14} + \pi_{23}) - \pi_{24}^2) -
\pi_{34}^2 = 0$$ in $G(2,4)$ for $J|_{{\cal V}_0} \in {\bf C}$.  Since
$G(2,4)$ is the vanishing set of the polynomial
\begin{equation} \pi_{12} \pi_{34} - \pi_{13} \pi_{24} + \pi_{14}
\pi_{23} = 0 \label{G24} \end{equation}
in $P(\wedge^2 {\bf C}^4)$, this base locus consists of two components,
$\pi_{34} = \pi_{24} = \pi_{14} = 0$ and $\pi_{34} = \pi_{24} =
\pi_{23} = 0$.  It is easily seen that each of these components is the
closure of a single torus orbit of complex dimension two.  Taking
their images under the momentum mapping, we obtain a statement
similar to Proposition~\ref{BIr1} for the base locus ${\cal
Z}_{0}$.   

\begin{prop}
$\tilde{\cal Z}_{0} = \mu_{G(2,4)}^{-1}(Star(\Box, L_1 + L_2))$.  Each of
the two components of $\tilde{\cal Z}_{0}$ is the closure of a single
torus orbit of complex 
dimension two; their images under the momentum mapping are the two
faces of $Star(\Box, L_1 + L_2)$.  
\end{prop}

\noindent This is illustrated in Figure 7.

%\begin{figure}[p]
%\vspace{18cm}
%\end{figure}

%\begin{figure}[p]
%\vspace{18cm}
%\end{figure}

%\begin{figure}[p]
%\vspace{18cm}
%\end{figure}

%\begin{figure}[p]
%\vspace{18cm}
%\end{figure}

\bibliographystyle{plain}
\bibliography{refs}

\end{document}